\documentstyle[pre,aps,multicol,epsf]{revtex}

\begin{document}
\draft

\title{A Model Ground State of Polyampholytes}
\author{Shay Wolfling and Yacov Kantor}
\address{School of Physics and Astronomy, Tel Aviv University,
Tel Aviv 69978, Israel}
\maketitle

\begin{abstract}
The ground state of randomly charged polyampholytes is conjectured to have
a structure similar to a necklace, made of weakly
charged parts of the chain, compacting into globules, connected
by highly charged stretched `strings'.
We suggest a specific structure, within the
necklace model, where all the neutral parts of the chain 
compact into globules:
The longest neutral segment compacts into a globule;
in the remaining part of the chain, the longest neutral segment
(the 2nd longest neutral segment) compacts into a globule, then the
3rd, and so on.
We investigate the size distributions of the longest neutral segments
in random charge sequences, using analytical and Monte Carlo methods. 
We show that the length of the $n$th longest neutral segment
in a sequence of $N$ monomers is proportional to $N/n^2$, while the
mean number of neutral segments increases as $\sqrt N$.
The polyampholyte in the ground state within our model
is found to have an average linear size proportional to $\sqrt N$, 
and an average surface area proportional to $N^{2/3}$.

\end{abstract}


\begin{multicols}{2}
\narrowtext
\section{Introduction}
\label{intro}
The desire to understand long chain biological macromolecules,
and especially proteins, stimulates extensive studies
of polymers \cite{CRE,GEN,GRO}. An important class of polymers are those 
with an electric charge along their backbone.
This work deals with heteropolymers, which carry positive and
negative charges, known as {\it polyampholytes} (PA's)
\cite{TAN}.
Models of PA's are important to the study of proteins, since
under normal physiological conditions, 5 of the 20 naturally occurring amino
acids have an excess charge \cite{CRE}.
We consider a polymer of charged monomers, interacting via unscreened Coulomb
interactions, and we investigate its ground state structure.
Throughout this work, we discuss PA's that 
consist of a random mixture of positive and negative charges, 
which cannot move along the chain.

We are interested in the geometrical features of the ground state of a PA,
and in particular in the dependence of its radius of gyration (r.m.s. size)
$R_g$ on the number of monomers $N$: $R_g\sim N^\nu$.
Higgs and Joanny \cite{HIG} and later Wittmer {\it et al.} \cite{WIT},
elaborated on arguments of Edwards {\it et al.} \cite{EDW} and
suggested, on the basis of a Debye-H\"uckel type theory \cite{LAN},
a collapsed structure ($\nu=1/3$) for neutral PA's in the ground state:
The chain takes advantage of the presence of two types of charges
along its backbone, and assumes a spatial conformation in which
every charge is predominantly surrounded by charges of an
opposite sign.

A different approach to the study of the ground state of PA's \cite{KK2},
is by scaling arguments, requiring that the interaction energy be
the same on all length scales. 
These arguments lead to a stretched structure of the ground state, 
where $R_g\sim N$.
In this approach $R_g$ is averaged over a complete ensemble of all
quenches, and the typical overall (excess) charge $Q$ of the PA
is $\sim \sqrt N$.
Whereas in the Debye-H\"uckel approximation, all the chains are
neutral, since the overall neutrality of the PA is an
essential condition for the validity of the Debye-H\"uckel approximation.
The extreme sensitivity of the ground state structure to the
excess charge, noted by Kantor {\it et al.} \cite{KLK,KKL},
and supported by Monte Carlo (MC) 
simulations as well as variational mean field calculations \cite{BRA},
resolves the apparent contradiction between the scaling and the 
Debye-H\"uckel motivated arguments.

To gain some insight into the behavior of PA's, 
analogies to charged drops were explored \cite{KK3,KK4,KK5}:
A spherical drop charged beyond a certain charge, called the 
Rayleigh charge $Q_R$, which depends on the surface tension and volume
of the drop, becomes locally unstable to elongation,
since the pressure difference between the inside and outside of the drop
vanishes.
Even before the total charge reaches $Q_R$, the drop becomes unstable to
splitting into two equal drops. Additional splittings of the
drop occur for larger charges. Similar behavior is expected in
PA's charged to $Q_R\sim\sqrt N$. 
Although a PA cannot split, the analogy to charged drops can still
be exploited:
Constraining the structure to maintain its
connectivity by attaching droplets with narrow tubes, results in a 
necklace-type structure of droplets connected by strings.

For homogeneously charged polymers (polyelectrolytes), the
charged drop analogy was used \cite{DRO} to characterize the
structure completely within the necklace model
(including the number of `beads' and strings, their sizes, and the 
number of monomers in them, for a given temperature and charge density).
However, trying to apply the necklace model to quenched PA's having 
random charges \cite{KK3,KK4}, 
several difficulties occur due to the randomness.
It was noted, for instance, that a situation occurs, in which most
spherical shapes are unstable, while there is on average no energetic
gain in splitting the sphere into two equal parts. A consistent theoretical
picture for random PA's beyond the instability threshold was not
found, but a typical PA is conjectured to be composed
of rather compact globules connected by long strings. In order
to reduce the electrostatical energy, the globules consist of
segments of the chain that are approximately neutral (collapsing
according to the Debye-H\"uckel picture), while the
strings are formed by highly charged segments.

Since the structure of a randomly charged PA cannot be characterized
completely analytically, we turn to
numerical MC methods in order to characterize the ground state of
such PA's within the necklace model.
A key role in the structure of randomly charged PA's is
played by the neutral segments in the chain
(forming the beads in the necklace).
We therefore apply MC methods to study the neutral segments
size distribution of randomly charged PA's.
The rest of the paper is organized as follows: 
In section \ref{model} we describe the process of construction of
a ground state for a randomly charged PA, by dividing it into
neutral segments. We discuss the motivation for this process, define
the important parameters of the problem and compare the process to 
similar existing models.
In section \ref{size} we investigate the sizes of the globules in
the suggested ground state, including their dependence on the total number
of monomers and on other parameters. 
In section \ref{finite} we discuss effects of finite chains, 
investigate the dependence of the total number of neutral segments 
in such finite chains on $N$, and construct a self consistent picture 
of the structure of the ground state. 
In section \ref{physical} we obtain some of the physical 
characteristics of the constructed ground state, such as its linear 
size and surface area, and in the last section we discuss our
results and compare them to other studies.

\section{The Model - Motivation and Definitions}
\label{model}
Empirical observations of the energy of randomly charged PA's 
suggested that the quench-averaged energy can be presented as a
sum of condensation, surface and electrostatic energies \cite{KK3,KK4}:
\begin{equation} 
\label{enerQ}
E=-\frac{q_0^2}{a}N+\gamma S + \frac{Q^2}{R}\ ,
\end{equation}  
where $q_0^2/a$ is the condensation energy gain per particle
($q_0$ is the typical charge of a monomer and $a$ is a microscopic
distance such as diameter of the monomer),
$\gamma\sim q_0^2/a^3$ is the surface tension, and $R$ is the
linear size of the chain. (In Eq.~(\ref{enerQ}) we omitted 
dimensionless prefactors of order unity.)
When the excess charge $Q$ is very small, the PA will form a
single globule. When $Q$ increases, the electrostatic forces will
overcome the surface tension and the globule will split
(similarly to the splitting of a drop charged beyond the Rayleigh
charge). Numerical studies suggest \cite{KK3,KK4,KK5} that a PA
forms a {\it necklace} of weakly charged globules, connected by highly
charged strings. This structure is a compromise between the tendency
to reduce the surface area (i.e. to form globules) due to surface tension,
and the tendency to expand, in order to reduce the Coulomb interaction
caused by the excess charge.

Kantor and Erta\c{s} \cite{KE1,KE2,KE3} attempted to quantify the
qualitative necklace model, by postulating that the ground state of
a PA will consist of a single globule, formed by the longest
neutral segment of the PA, while the remaining part will form a tail.
They investigated the probability that the longest neutral
segment in a chain of $N$ monomers has length $L$. 
A probability density was defined 
and investigated in the limit where $N,L\rightarrow\infty$, 
while the reduced length $l\equiv L/N$ is fixed.
The probability density was obtained numerically and investigated 
analytically, but a complete analytical solution to the problem 
was not found.
We follow \cite{KE1}, and investigate the problem of the size 
distribution of neutral segments in randomly charged PA's 
by mapping the charge sequence into a one-dimensional random walk (1-d RW).
The charge sequence $\omega=\{q_i\}\ (i=1,...,N;\ q_i=\pm1)$ is mapped 
into a sequence of positions $S_i(\omega)=\sum_{j=1}^i q_j$ ($S_0=0$)
of a random walker. (From now we will measure charges in units of the
basic charge $q_0$, and therefore $q_j$ will be dimensionless.)
The random sequence of charges is thus equivalent to a RW,
a chain segment with an excess charge $Q$ corresponds to a
RW segment with total displacement of $Q$ steps, and
a neutral segment is equivalent to a loop inside the RW.
Throughout the paper we will use the terminologies
of randomly charged PA's and of RW's interchangeably.

\begin{figure}[]
\centerline{\hbox{
\epsfysize=18\baselineskip
      \epsffile{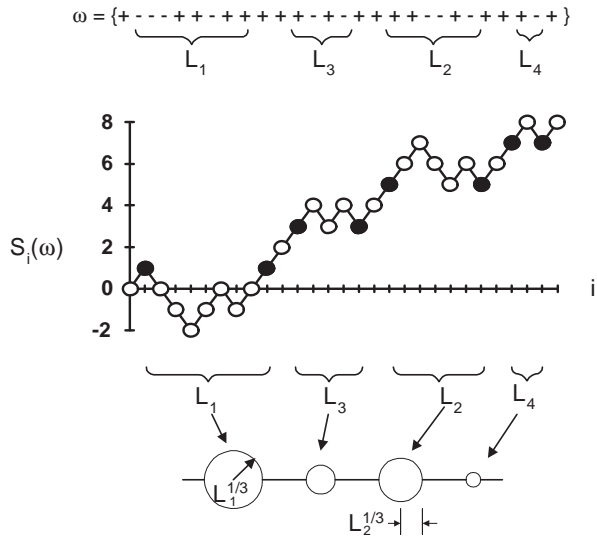}  }}
\caption [Figure]
	 {\protect\footnotesize  An example of a charge sequence $\omega$, 
	 mapped into a 1-d RW $S_i(\omega)$, and a typical loops structure.
	 Filled circles indicate the starting and ending points of loops.
	 The longest loop in the RW has 8 steps ($L_1=8$), the 2nd
	 longest loop has 6 steps ($L_2=6$), $L_3=4$ and $L_4=2$.
	 The excess charge (which is equivalent to the total displacement
	 of the RW) is $Q=+8$ and the total length is $N=28$.}  
\label{figure1}
\end{figure}
When the longest neutral segment forms a globule (as assumed in 
\cite{KE1,KE2,KE3}), the remaining part
of the chain is very large ($\sim N$). It is natural to assume that
neutral segments on that tail will further reduce the total energy
by folding into globules. Eventually, the necklace will consist of many
neutral globules. However, there are many ways in which the chain can
be divided into neutral segments, and we are interested in a simple
unique structure.
We therefore suggest a {\it specific} necklace-type structure, and construct 
the ground state for a randomly charged PA in the following way, depicted
in Fig.\,\ref{figure1}: 
The longest neutral segment contains $L_1$ monomers; it compacts into a
globule of linear size proportional to $L_1^{1/3}$; in
the remaining part of the chain the longest neutral segment (the 2nd longest
neutral segment) of size $L_2$ also compacts into a globule, 
then the 3rd and so on, until the segments become very small 
(of only a few monomers). Generally, $L_n$ denotes the number of monomers
in the $n$th longest neutral segment, which compacts into a globule of radius
$L_n^{1/3}$. Eventually, all the neutral segments are exhausted and we 
are left only with strings which carry the PA's excess 
charge $Q$, and connect the globules. The total number of monomers in
the chain is the number of monomers in all the neutral segments, 
plus the excess charge.

The `ground state', generated by this process of 
longest neutral segments compacting into globules, does not necessarily 
minimize the total energy of the PA. 
First of all, the process does not consider
the possibility of weakly charged globules, which can include much 
more monomers than the neutral globules, thus compensating by surface 
energy for the additional electrostatic energy. 
Secondly, even when considering 
only neutral globules, it is not necessary that the procedure of
compacting the longest neutral segment at each step generates the 
lowest energy state: It is possible that two `medium sized' globules
will have lower surface energy than a long one and a very short one (which
remains after the first long segment was already chosen).

A similar model, describing a breaking of a RW into loops, is the
Loop Erased Self-Avoiding Walk (LESAW) model \cite{LA1}.
A LESAW is constructed from a RW, in which any loop generated by a self
intersection is erased. 
The LESAW model was subject for intensive study \cite{LA2,LA3,GTM,BRD,DHA},
and many properties of the lengths of the erased loops in it
were found.
However, these results do not consider issues of the {\it longest}
erased loops, and are therefore of little help to us.

Other related models are those of randomly broken objects
\cite {DR1,DR2}, in which an object, such as
a segment of unit length, is divided into
mutually exclusive parts by a self-similar random process.
The division of a segment into parts resembles the
`breaking' of a RW into loops: The probability densities of the longest 
segments in some models of randomly broken objects resemble
the probability density of the longest loop in \cite{KE2}.
However, one main difference between models of randomly broken objects 
and the longest loop problem, is that the probabilities of the longest
loops are not self-similar. (The probability of a certain
fraction of the chain to form the longest loop and the probability of 
the same fraction of the remaining chain to form a
second longest loop are different.)
We will return to this absence of self-similarity in section
\ref{size}.

\section{Size Distribution of Neutral Segments}
\label{size}
In order to characterize a PA in the necklace-type low energy state,
made of neutral globules and highly charged strings, 
we are interested not just in the size distribution of the longest 
neutral segment, but also in the size distribution of the 2nd 
longest segment, 3rd longest, and so on.
We examined the statistics of loops in a 1-d RW of 
$N$ steps, using MC method for several
$N$'s, up to $N=10^4$. For each $N$ we randomly selected
$10^6$ sequences. For each sequence we found the lengths $L_n$ of the 
(non-overlapping) loops, and calculated their distributions and averages.

We denote by $P_{N,n}(L_n)$ the probability of the $n$th longest loop in a 
RW of $N$ steps to be of length $L_n$. 
The average length of the longest loop was found \cite{KE1,KE2} 
to be proportional to $N$. Expecting the same behavior for the
length of the $n$th longest loop, we define the probability density 
of the $n$th longest loop:
\begin{equation}
\label{pdef}
p_n(l_n)\equiv\frac{N}{2}\left[P_{N,n}(L_n)+P_{N,n}(L_n+1)\right] \ ,
\end{equation}
where $l_n\equiv L_n/N$. 
(Note that at least one term in Eq.~(\ref{pdef}) vanishes, since loops
can be only of even length. Therefore, definition (\ref{pdef})
includes average of probabilities for $L_n$ and $L_n+1$ as in the
definitions used in continuum limits for discrete RW's, in order
to prevent even-odd oscillations.)
For small $N$'s this definition of $p_n(l_n)$ still
depends on $N$, but in the $N,L_n\rightarrow\infty$ limit it
becomes a function of only $l_n$.
Numerical evidence for the $N-$independence of $p_1(l_1)$ 
for $N\rightarrow\infty$ was presented in \cite{KE2}. 
MC results for $p_n(l_n)$ for several values of $n$ indicate that 
the probability densities are virtually independent of $N$ \cite{FUT}.
For small $N$ and large $n$ the probability density $p_n(l_n)$
depends on $N$, since $L_n$'s are short, and continuum limit is 
expected only when $L_n\gg1$. 
In order to overcome effects of finite $N$, we determined the
$N-$dependence of $\langle L_n\rangle$, through an extrapolation of the
slopes of the linear fits of $\log\langle L_n\rangle$ {\it vs.} $\log N$ 
to $N\rightarrow\infty$. For any given $n$
we calculated, through a local linear fit, slopes for several values of $N$,
and estimated them at $N\rightarrow\infty$ limit.
Furthermore, it can be proven analytically \cite{FUT}
that the probability density of the longest loop is `universal', i.e.
for long RW's, it does not depend on the number of steps, or on details
of a single step.

\begin{figure}[]
\centerline{\hbox{
\epsfysize=18\baselineskip
      \epsffile{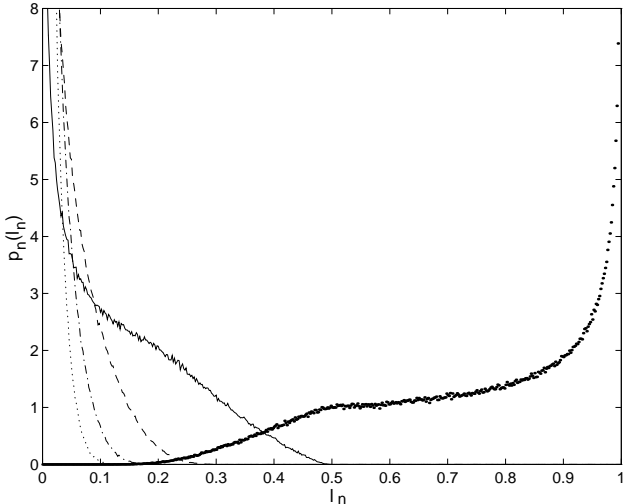}  }}
\caption [Figure.]
	 {\protect\footnotesize  Probability densities of the 5 longest loops
 from right to left: $p_1(l_1)$-thick points, $p_2(l_2)$-solid line, 
 $p_3(l_3)$-dashed line, $p_4(l_4)$-dot-dashed line, $p_5(l_5)$-dotted line, 
 as a function of $l_n=L_n/N$ from MC
 results of $10^6$ random sequences of length $N=1000$.}  
\label{figure2}
\end{figure}

Since $p_n(l_n)$ are $N-$independent, we can investigate them
for any given (large enough) $N$. The probability densities of the five
longest loops in a chain are depicted in Fig.\,\ref{figure2} for $N=1000$.
Several properties of $p_n(l_n)$ are evident from this figure.
The probability density $p_1(l_1)$ was shown
\cite{KE2} to have a square root divergence at $l_1=1$, 
and a discontinuous derivative at $l_1=\frac{1}{2}$.
When the length of the first loop $L_1$ in a specific chain is long,
then the lengths of the other loops $L_n$ (for $n>1$) in that chain must
be short, since the total length of all the loops cannot exceed $N$.
Therefore, since in many chains the length of the longest loop is almost
equal to $N$ (as indicated by the divergence of $p_1(l_1)$
when $l_1\rightarrow1$), the lengths of the other loops approach
zero. This is evident from the divergence of $p_n(l_n)$
when $l_n\rightarrow0$ for all $n>1$.
Because in any specific chain the length of the
$n$th longest loop is shorter than the length of the $k$th
longest loop, for $n>k$, the divergence of $p_n(l_n)$
near zero is stronger for large $n$.
Since the probability densities $p_n(l_n)$ are
normalized separately for each $n$, then any two of them
must intersect (i.e. $p_n(l_n)$ always intersects $p_{n'}(l_{n'})$
for $n\not=n'$).
The length of the second longest loop never exceeds the length
of the first longest loop, and the sum of their lengths never exceeds $N$.
Therefore, the length of the second longest loop cannot exceed half
the length of the chain. Consequently, $l_2\le\frac{1}{2}$ 
for all the chains, and $p_2(l_2)$ vanishes identically for $l_2>\frac{1}{2}$.
Similarly we can show that $l_n\le 1/n$ for all $n$ and $p_n(l_n)=0$ 
for $l_n>1/n$.

We note that $p_2(l_2)$ is qualitatively
similar to the probability densities of the length of the second segment in 
different models of randomly broken objects \cite{DR1,DR2}
(see Figs. 1(b) and 3(b) in \cite{DR1}).
However, the probability densities of the length of the
second segment in \cite{DR1} have strong singularities (of the
first derivative) at $l_2=1/4$, and are shown to have
singularities at $l_2=1/k$ for all integer $k\ge 2$.
All the probability densities $p_n(l_n)$ of the longest loops
have a singularity when they become identically zero
($p_n(l_n)$ vanishes for $l_n>1/n$ and therefore is non-analytical
at $l_n=1/n$).
It is possible that due to the singularity in $p_{n'} (l_{n'})$ at $1/n'$
all the other probability densities $p_n(l_n)$ (for $n<n'$) also
have singularities at $1/n'$, since all the probability densities are
dependent.
Apparently, these singularities do not cause discontinuity
of the first derivative, and, therefore are not visible in the numeric data.
In several models of randomly broken objects \cite{DR1,DR2},
the probability densities of the length of the $n$th segment have 
singularities at $l=1/n'$ (for $n\le n'$). These singularities are due 
to a self-similar process, which leads to a different analytical expression 
for the probabilities on each interval $1/(n'+1)<l<1/n'$.
Since the random process of generating the longest loops in our
process is not self-similar (as indicated in the previous section),
the reason for the singularities in the models of randomly broken objects
does not hold in
the case of the $n$th longest loop (as was suggested by
Frachebourg {\it et al.} \cite{FRA}). Therefore, the probability 
of the $n$th longest loop does not necessarily have singularities
at values of $l=1/n'$.

\begin{figure}[]
\centerline{\hbox{
\epsfysize=16\baselineskip
      \epsffile{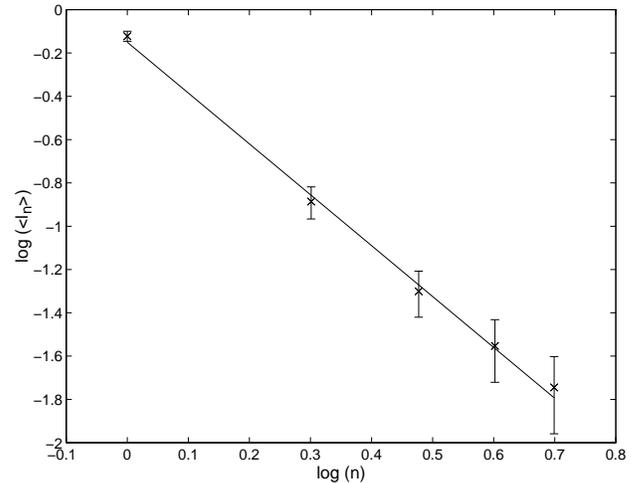}  }}
\caption [Figure.]
	 {\protect\footnotesize Logarithm (base 10) of average reduced length
	 of the $n$th longest loop as a function of logarithm of
	 loop number $n$.}
\label{figure3}
\end{figure}

Since in any specific RW, $L_1\ge L_2\ge \cdots \ge L_n$, the 
average length of the $n$th longest loop $\langle L_n\rangle$ decreases 
(for fixed $N$) with increasing loop number $n$. 
There is no typical scale in the problem, and we may expect a power law
dependence $\langle L_n\rangle\sim N n^{-\alpha}$, with $\alpha>0$.
The total number of steps in all the loops in any given RW cannot exceed
$N$ (i.e. $\sum_n L_n\le N$), and therefore
$\sum_n \langle L_n\rangle\le N$.
The convergence of the sum means that $\alpha>1$.
We depict in Fig.\,\ref{figure3} the average reduced length 
$\langle l_n\rangle$ {\it vs.} $n$ on a logarithmic scale.
To avoid systematic errors due to finite $N$, each value of
$\langle l_n\rangle$ in the graph was determined through an extrapolation:
For each $n$, we plotted $\langle l_n\rangle$ {\it vs.} $1/N$, and
found the extrapolated value of $\langle l_n\rangle$ near $1/N=0$.
These values of $\langle l_n\rangle$ are depicted in Fig.\,\ref{figure3}.
The linear fit to the data points has a slope of
$-2.3\pm0.4$. We therefore conclude that as $N\rightarrow\infty$:
\begin{equation} 
\label{ldp}
\left<L_n\right>\sim \frac{N}{n^\alpha}\ \ ,\ \ 
{\rm where}\ \ \alpha = 2.3\pm0.4\ .
\end{equation}  
At the next section we will argue that
$\alpha=2$, which is within the error limits of Eq.~(\ref{ldp}).

\section{Neutral Segments in Finite Chains}
\label{finite}
So far we have discussed only long chains and their
properties in the $N\rightarrow\infty$ limit. 
There are several differences between the properties of infinitely
long chains and finite size chains.
First of all, in finite size chains the monomers that do not belong to
any neutral segment 
constitute a finite part of the chain, as opposed to a vanishing
part as $N\rightarrow\infty$. Secondly, in infinitely long chains
the number of neutral segments is infinite, while for finite $N$
at some point there are no more neutral segments.

Let us consider the total number of monomers
which do not belong to any neutral segment, for finite $N$.
It is easily shown that the number of monomers left out from all 
the neutral segments in any specific random sequence is exactly the 
excess charge $Q$ of the corresponding PA:
The excess charge cannot exceed the number of monomers left out,
because, by definition, the neutral segments do not have excess charge.
On the other hand, all the monomers left out from the neutral segments
must be part of the excess charge: If, for instance, a PA has
excess positive charge, then no negative charge can be left out from all
the neutral segments. (If there had been a group of negative charges, 
it would have joined a group of positive charges to make a neutral segment,
or together with a group of positive charges, would have joined an existing 
neutral segment and make it longer).
Since the r.m.s. excess charge of a randomly 
charged sequence of $N$ charges of $\pm1$ is equal to $\sqrt N$,
the r.m.s. number of monomers not in any neutral segment in the chain 
is also equal to $\sqrt N$, becoming a vanishing fraction of the chain 
as $N\rightarrow\infty$.

At some point the process of search for the next longest loop exhausts
all the loops in the RW.
We investigate this stage in the process, by analyzing 
the number $n_f$ of loops in a RW.
When $N$ is infinite, the average length of the $n$th longest loop
is given by $\langle L_n\rangle\sim Nn^{-\alpha}$.
Application of this equality for finite $N$'s would predict, for large 
enough $n$'s, $\langle L_n\rangle\le1$.
Since this is not possible (the minimal length of a loop is two steps),
we argue that Eq.~(\ref{ldp}) is valid, for finite $N$'s, only
to describe the average lengths of the longest $n_f$ loops.
There is no typical scale to the problem of the total number of loops,
and we therefore expect a power
law dependence of $\langle n_f\rangle$ on $N$. 
The length of the last loop, for all chain lengths, is usually 
very short (consisting of only few positive and negative charges), 
having length independent of $N$,
i.e. $\langle L_{n_f}\rangle\sim N^0$, which means that
$\langle l_{n_f}\rangle\sim N^{-1}$. Since  
$\langle l_{n_f}\rangle\sim n_f^{-\alpha}$ (substituting $n=n_f$ in
Eq.~\ref{ldp}), we can expect:  
\begin{equation}  
\label{y1al}
\langle n_f\rangle\sim N^y\ ,\ \ \ {\rm where}\ \ y = \frac{1}{\alpha} \ .
\end{equation} 
Substituting $\alpha$ from Eq.~(\ref{ldp}), leads to a value of 
$y=0.43\pm0.09$.
Investigating numerically the dependence of $\langle n_f\rangle$ on $N$, 
we found that $y=0.46\pm0.06$,
which is within the error limits of the value
predicted by Eqs.~(\ref{ldp}) and (\ref{y1al}).

\begin{figure}[]
\centerline{\hbox{
\epsfysize=16\baselineskip
      \epsffile{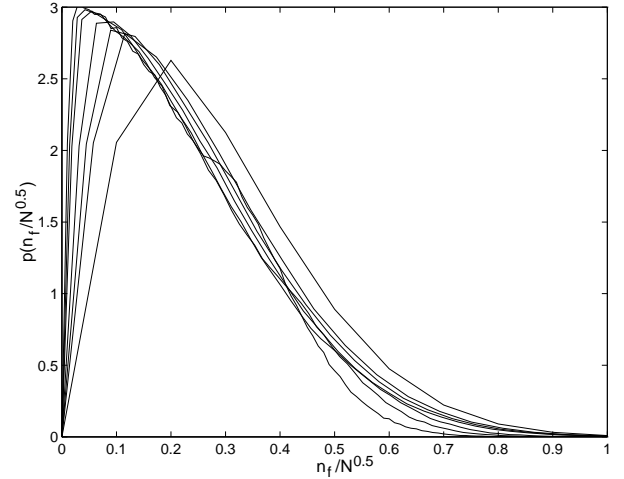}  }}
\caption [Figure.]
	 {\protect\footnotesize Probability density of $n_f$, the total 
	 number of loops in a chain, divided by $\sqrt N$, for  
	 $N=$ 100, 300, 500, 1000, 3000, 5000, 10000 (from right to
	 left).}  
\label{figure4}
\end{figure}

At the end of this section we will argue that $\alpha=2$ and 
$y=0.5$, values which are within the error limits of those deduced from
the numeric data.
In order to confirm the $\langle n_f\rangle \sim \sqrt N$ relation,  
we show in Fig.\,\ref{figure4}
the probability density of $n_f$ divided by $\sqrt N$
for several chain lengths ($N=10^2\div10^4$). 
The division by $\sqrt N$ causes a reasonable collapse of the graphs for
different values of $N$ to a single function, which is (almost)
$N-$independent. 
We see that the probability density has a maximum 
when $n_f/\sqrt N$ is close to zero
(the most probable value of $n_f$ is finite and independent of $N$), and it
decreases to zero with increasing $n_f/\sqrt N$.

Knowing the statistical properties of the chain, we try to
construct a self-consistent complete picture, in which our numerical
results fit together. We know that the total length of all the loops 
($\sum_{n=1}^{n_f}L_n$) equals to the entire length of the chain minus 
the steps not in any
loop (which are the excess charge, which are on the average $\sqrt N$). 
Dividing this equality by $N$, taking an average over the random sequences, 
we get:
\begin{equation}
\label{four7}
1-\frac{1}{N^{\frac{1}{2}}} = \langle \sum_{n=1}^{n_f}l_n\rangle\ .
\end{equation}
On the other hand, from Eqs.~(\ref{ldp}) and (\ref{y1al}) we know that,
omitting constants of order unity:
\begin{equation}
\label{four8}
\left<\sum_{n=1}^{\langle n_f\rangle}l_n\right>=
\sum_{n=1}^{\langle n_f\rangle}\left<l_n\right>\simeq
\int_{n=1}^{N^y}n^{-\alpha}dn\sim 1-N^{y(1-\alpha)}\ .
\end{equation}
Comparison of the powers of $N$ in Eqs.~(\ref{four7}) and (\ref{four8}) 
leads to
\begin{equation}
\label{condy1}
y = \frac{1}{2(\alpha-1)}\ ,
\end{equation}
which together with Eq.~(\ref{y1al}) is satisfied by:
\begin{eqnarray}
\label{yandal}
\alpha=2\ ,\ \ \ \ \  y=\frac{1}{2}\ .
\end{eqnarray}
These equalities are satisfied by the values of $y=0.46\pm0.06$,
and $\alpha=2.3\pm0.4$ obtained numerically, and
constitute a self-consistent picture, in which the 
average conformational properties of the constructed ground state
fit together.

\section{Physical Properties of the Ground State}
\label{physical}
In this section we investigate some of the physical characteristics
of the constructed ground state of randomly charged PA's.
We focus on the linear size $R$ and on the surface area $S$ of the
proposed ground state, 
trying to explain their dependencies on $N$ through the self
consistent picture constructed in the previous section.
We define $R$, the linear size of the chain, according to the picture
of Fig.\,\ref{figure1}: The neutral segments in the chain compact into
globules (each with a linear size of 
$R_{\rm segment}\sim L_{\rm segment}^{1/3}$).
If all the globules are linearly packed then the total linear size
is the sum of the linear sizes of all the globules
($R=\sum R_{\rm segment}$).
The linear size of the chain must include the monomers not
in any neutral segment (the `strings' in the necklace, which are the
total excess charge $Q$).
We therefore get a means to describe
the chain's size:
\begin{equation}
\label{rdf}
R\equiv\sum_{n=1}^{n_f} L_n^{1/3} + Q \ .
\end{equation}
(Here, and throughout this section, we omit prefactors of order unity.)
It is evident that the generated state captures some essential 
features of the  ground state suggested by the necklace model: The 
necklace type structure is compact (i.e. $R\sim N^{1/3}$) when
the PA is neutral (the longest neutral segment is the entire chain) or has
very small excess charge, and begins to stretch as the excess charge
increases (the charged strings become longer). Finally, the PA becomes 
completely stretched (i.e. $R\sim N$) for the fully charged polymer.

\begin{figure}[]
\centerline{\hbox{
\epsfysize=15\baselineskip
      \epsffile{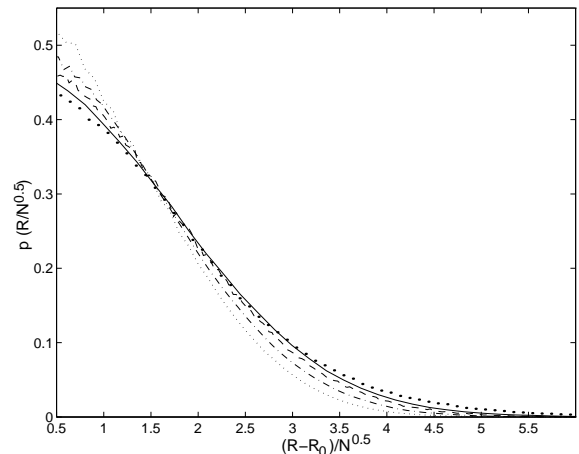}  }}
\caption [Figure.]
	 {\protect\footnotesize Probability density of $R$, 
	 the linear size of the  entire chain, for several chain 
	 lengths: 
	 $N=$100 (dotted line), 300 (dot-dashed line), 1000 (dashed line), 
	 3000 (solid line), 10000 (thick points).
	 The minimal possible value of $R$ ($R_0=N^{1/3}$) is subtracted
	 from $R$, and the result is divided by $\sqrt N$ to collapse 
	 the data.}  
\label{figure5}
\end{figure}

Investigating the average chain's linear size dependence on $N$,
we get the dependence:
\begin{equation} 
\label{rnu}
\left<R\right>\sim N^\nu\ ,\ \ {\rm where}\ \  \nu=0.50\pm0.01\ .
\end{equation}  
Fig.\,\ref{figure5} depicts the probability density of $R$, divided by 
$\sqrt N$ for several values of $N$. 
From the data collapse we deduce that the $N-$dependence 
in Eq.~(\ref{rnu}) is valid not just for the average linear size,
but represents a scaling of the entire probability density.
The $\nu\simeq\frac{1}{2}$ power in Eq.~(\ref{rnu}) means that the chain is 
not compact (although the distribution is `peaked' near the lowest 
possible value of $R$), and is not completely stretched, but has a 
linear size as an ideal RW with $N$ steps.

We can explain the dependence in Eq.~(\ref{rnu}), by assuming
that $\sum_n\langle {L_n}^{1/3}\rangle$ and
$\sum_n {\langle L_n\rangle}^{1/3}$ have the same $N-$dependence,
and by using the power laws of $\langle n_f\rangle$ and 
$\langle L_n\rangle$ (Eqs.~\ref{ldp} and \ref{y1al}), with
$\alpha=2$ and $y=\frac{1}{2}$ (Eq.~\ref{yandal}):
\begin{equation}
\label{Rtil}
\sum_{n=1}^{n_f}\langle L_n^{1/3}\rangle \sim
\sum_{n=1}^{\langle n_f\rangle}\langle{L_n}\rangle^{1/3} \sim
\int_{n=1}^{N^y}{\left(\frac{N}{n^\alpha}\right)}^{1/3} \sim
N^{0.5}\ .
\end{equation}
In order to confirm this dependence, we investigated
$\langle\sum_{n=1}^{n_f} L_n^{1/3}\rangle$ as a function of $N$,
and found the power law dependence with exponent
$0.49\pm0.02$, in accordance with the prediction of Eq.~(\ref{Rtil}).
We see that the average linear chain size $\langle R\rangle$ in 
Eq.~(\ref{rdf}) is a sum of two terms, each proportional to $\sqrt N$.

Similarly to the definition of $R$, we can define the surface area $S$
of the chain. Each neutral segment compacts into a globule of surface
area $\sim L_{\rm segment }^{2/3}$, all the globules are linearly packed,
and the surface area of the `strings' is proportional to the number of 
monomers not in any neutral segment, which is the excess charge. 
We can therefore define:
\begin{equation}
\label{sdf}
S\equiv\sum_{n=1}^{n_f} L_n^{2/3} + Q\ .
\end{equation}
$S$ can have values ranging from $N^{2/3}$, for a neutral chain, to $N$, 
for a completely charged chain.
These limits indeed correspond to the expected behavior of PA.
Obviously $S\ge N^{2/3}$.
We analyzed the $N$-dependence of $S$, and found that:
\begin{equation}
\label{four19}
\langle S\rangle
\sim N^{0.67\pm0.01}\ .
\end{equation}
When we subtract from $S$ its minimal value, and divide the result by
$N^{2/3}$, we get a distribution which is identical for all $N$.
This dependence means that the average surface energy of the generated 
structure has the same $N-$dependence as 
the surface energy of a single compact globule (or several compact globules,
each containing a finite part of the chain). 
From the same arguments that led to Eq.~(\ref{Rtil}), we get:
\begin{equation}
\langle S\rangle = 
\langle Q\rangle + \sum_{n=1}^{\langle n_f\rangle}\langle {L_n}\rangle^{2/3} 
\sim \sqrt{N} +\!\! \int_{n=1}^{N^y}{\left(\frac{N}{n^\alpha}\right)}^{2/3} 
\!\!\!\!\sim N^{2/3}\ .
\end{equation}
This power of $N$ is in accordance with the numerically obtained
value of $0.67\pm0.01$, and indicates 
that the $N$-dependence of the average surface area is determined by the
neutral segments (i.e. the `beads' in the necklace), and is not
affected by the excess charge (the `strings' in the necklace).

We investigated the expression $S+\frac{Q^2}{R}$, which has the same 
$N$-dependence as the energy of the generated structure. (We considered
the energy terms of Eq.~(\ref{enerQ}), and omitted the condensation term,
since it is the same for all structures of a given $N$).
Exploring the $N-$dependence of $S+\frac{Q^2}{R}$ (denoted as $E$),
we found that:
\begin{equation} 
\label{edef} 
\langle E\rangle\equiv\langle S+\frac{Q^2}{R}\rangle\sim N^{0.66\pm0.01} \ .
\end{equation}  
When subtracting from $E$ its minimal value and dividing the result 
by $N^{2/3}$, we get a probability density which is identical for all $N$.
This dependence means that the total energy is very low:
The surface energy term ($S$) gets its minimal value ($\sim N^{2/3}$), and 
$R\sim N^{1/2}$, thus bringing the electrostatic energy term to a point where
it does not affect the $N$-dependence of the total energy.
The total energy behaves very much like
the surface energy, as if the chain is constructed of a
{\it single compact neutral globule}.

\section{Conclusions and Discussion}
We have studied the properties of a conjectured structure of randomly
charged PA's in the ground state.
According to the necklace model \cite{KK3,KK4}, in the ground state of 
PA's neutral segments in the chain compact into globules. 
Following Kantor and Erta\c{s} \cite{KE1,KE2}, we have mapped 
the problem of longest neutral segments to the problem of longest loops
in 1-d RW's, and applied numerical methods along 
with analytical estimates to study the size distribution of longest loops.

Since we believe that a structure of PA's based on the necklace model
has a very low energy, we suggested a specific detailed necklace-type 
structure for polyampholytes in the
ground state, and numerically obtained its conformational and physical 
properties (the number and sizes of `beads' and `strings' 
in the necklace, the spatial extent, the surface area and energy). 
This structure is compact when the chain is neutral or
weakly charged, and stretches as the chain becomes charged.
We have found that the ground state structure has a very low energy, which
depends on the number of monomers as the energy of a single compact neutral
globule. 
We have also shown that the unrestricted average of the linear size 
of the polymer in the ground state depends on the number of monomers 
as the linear size of an ideal chain, with a critical exponent of
$\nu=0.50\pm0.01$. 
Although the qualitative behavior of our distribution for the linear
size is similar to this obtained in \cite{KK5}
(it is peaked near its smallest possible value, and
has a tail, which determines the asymptotic behavior of $\nu$),
Kantor and Kardar \cite{KK5} concluded that the average linear size increases
with $N$ at least as fast as a self-avoiding walk (i.e. $\nu>0.6$).
We believe that it is worthwhile to slightly alter the way
in which compact globules are formed within our model (by allowing for
weakly charged globules or by not forcing all the neutral segments to
completely compact), in order to try to reproduce this `swelling'
of the average chain.

The investigation of the size distribution of neutral segments through
the analogy to 1-d RW's, was limited to a particular class of RW's,
in which a unit displacement appears at each step (i.e. each monomer
in the chain is charged $\pm1$).
In a future publication \cite{FUT}, we define the problem of longest loops
in the limit where the RW becomes a true Gaussian walk, and prove 
through a scaling process the universality of the probability densities
of longest loops. This universality means that the probability
densities of longest loops are independent of the number of
steps and of the nature of the single step of the RW.
Therefore, the results obtained for a specific case of randomly charged
PA's are valid for all randomly charged polymers, in which 
(for long chains) the charge distribution is an unbiased Gaussian.

\section*{Acknowledgments}
This work was supported by the Israel Science Foundation under 
grant No. 246/96.


\end{multicols}
\end{document}